\documentclass[twocolumn,showpacs,preprintnumbers,amsmath,amssymb]{revtex4}

\usepackage{graphicx}
\usepackage{dcolumn}
\usepackage{bm}

\begin{document}

\title{The band structure of hexagonal diborides ZrB$_2$, VB$_2$, NbB$_2$ and TaB$_2$\\
in comparison with the superconducting MgB$_2$}

\author{I.R. Shein, A.L. Ivanovskii}

\address{Institute of Solid State Chemistry, Ural Branch of the Russian
Academy of Sciences, 620219 Ekaterinburg, Russia}
\email{shein@ihim.uran.ru}
\date{\today}

\begin{abstract}
The band structure  and the Fermi surface of hexagonal diborides
ZrB$_2$, VB$_2$, NbB$_2$, TaB$_2$ have been studied by the
self-consistent full-potential LMTO method and compared with those
for the isostructural superconductor MgB$_2$. Factors responsible
for the superconducting properties of AlB$_2$-like diborides are
analyzed, and the results obtained are compared with previous
calculations and available experimental data.
\end{abstract}


\pacs{71.20.Cf, 71.90+q, 74.25.Jb}
\maketitle A recent discovery [1] of the critical transition
(T$_c$ $\approx$ 40 K) in magnesium diboride (MgB$_2$) and
development of some promising superconducting materials on its
basis (ceramics, films, extended wires, see review [2]) gave an
impetus to active search for novel superconductors (SC) among
related compounds with the structure and chemistry similar to
MgB$_2$. It is reasonable that hexagonal (AlB$_2$-like) metal
diborides isostructural with MgB$_2$ were considered as first SC
candidates. Detailed investigations of the band structure and
coupling mechanism in MgB$_2$ [2-7] showed that diborides of group
I, II metals of the Periodic System, for instance metastable
CaB$_2$ [6], LiB$_2$, ZnB$_2$ [7], hold the greatest promise as SC
candidates. The authors of [8] predicted the possibility of
critical transition with T$_c$ $>$ 50 K in AgB$_2$ and AuB$_2$. It
is much less probable to find new SC (with T$_c$ $>$ 1 K) among
AlB$_2$-like diborides of d metals (MB2), see [2-7]. The first
report ([9], 1970) on the superconductivity in NbB$_2$ (T$_c$
$\approx$ 3.9 K) was not confirmed by systematic studies [10] of
SC properties of a series of diborides MgB$_2$ (M = Ti, Zr, Hf, V,
Nb, Ta, Cr), according to which their T$_c$ are smaller than 0.7
K.\\ Therefore recent reports [11-13] on rather high T$_c$ for
ZrB$_2$ (5.5 K [11]), TaB$_2$ (9.5 K [12]) and NbB$_2$ (5.2 K
[13]) were quite unexpected. It is remarkable that in the
investigations of identical series of diborides (TiB$_2$, ZrB$_2$,
HfB$_2$, VB$_2$, NbB$_2$, TaB$_2$ [12] and ZrB$_2$, NbB$_2$,
TaB$_2$ [11]) each group revealed "its" superconductor, namely
ZrB$_2$ [11] and TaB$_2$ [12], while all other MB$_2$ phases were
assigned to non-superconductors.\\The results published in [12]
prompted the authors of [14] to examine in detail the temperature
dependencies of magnetic susceptibility and electrical resistance
in TaB$_2$. It was established that the SC transition for TaB$_2$
is not observed to T $\approx$ 1.5 K. The SC properties of MgB$_2$
and TaB$_2$ discussed [14, 15] on the basis of the band structure
calculations were found to be essentially different due to strong
hybridization effects of the Ta5d-B2p states. A weak (as compared
with MgB$_2$) electronic interaction with the E$_2g$ mode of the
phonon spectrum was noted in [14]. An abrupt lowering of T$_c$ for
TaB$_2$ (and the absence of SC for VB$_2$) is explained in [15] by
a considerable decrease in the contributions from B2p states to
the density of states at the Fermi level (N(E$_F$)): MgB$_2$
(0.494) $>$ TaB$_2$ (0.114) $>$ VB$_2$ (0.043 states/eV).
Analyzing fine features of soft X-ray emission and absorption
BK-spectra of MgB$_2$, NbB$_2$ and TaB$_2$, the authors of [16]
pointed out fundamental differences in the structure of their
pre-Fermi edges with dominating contributions from B2p$_\sigma$
(MgB$_2$) or B2p$_\pi$ states (NbB$_2$, TaB$_2$). Any operations
reproducing results [11], to us are unknown. As is shown in [2-7],
the superconductivity in MgB$_2$ and related borides is fairly
well described in the context of the electron-phonon interaction
theory. Therefore the peculiarities of the electronic spectrum,
primarily the composition and structure of the near-Fermi bands,
constitute the major factor responsible for the formation of this
effect. In this work we present the results of detailed band
structure analysis of Zr, V, Nb and Ta diborides in comparison
with the superconducting MgB$_2$. As is known, these diborides are
isostructural (AlB$_2$ type, space group P6/mmm), their crystal
lattices are made up of alternating hexagonal metal (M) monolayers
and graphite-like boron layers [17]. The unit cell contains three
atoms (M, 2B). The main differences are due to the metal
sublattice type, viz. the electronic configurations of metal atoms
(Mg - 3s$^2$3p$^0$; Zr - 5s$^2$4d$^2$; V, Nb, Ta -
(n+1)s$^2$nd$^3$, where n = 3, 4, 5 respectively), which determine
the electron concentration (EC) growth (MgB$_2$ (8) $<$ ZrB$_2$
(10) $<$ VB$_2$, NbB$_2$, TaB$_2$ (11 e/cell)) and the variation
in interatomic bonds, see [2, 18, 19]. \\The band structure of
MgB$_2$ was calculated in the framework of the LDA theory by the
self-consistent full-potential linear muffin-tin orbital (FP LMTO)
method with allowance for relativistic effects and spin-orbital
interactions [20, 21] with the exchange-correlation potential in
accordance with [22]. Equilibrium parameters of MB$_2$ cells
(Table 1) were obtained based on the condition of the system's
total energy minimum.\\\textbf{MgB$_2$}. As follows from Figs. 1
and 2, the peculiarities of the band structure of the
superconducting MgB$_2$ are due to the B2p states, that form four
$\sigma$(2p$_{x,y}$) and two $\pi$(p$_z$) energy bands. The
B2p$_z$ states are perpendicular to boron atom layers and form
weak interlayer $\pi$ bonds. The B2p$_{x,y}$ bands are of
two-dimensional-like (2D) type and form flat areas in the
$\Gamma$-A direction of the Brillouin Zone (BZ). A small
dispersion of $\sigma$ bands is also indicative of an
insignificant interaction between Mg-B layers. Two B2p$_{x,y}$
bands intersect E$_F$ and make an appreciable contribution to the
density of states (DOS) at the Fermi level being responsible for
the metal-like properties of MgB$_2$,see Table 2. One of the most
important features of MgB$_2$ is the presence of hole B2p$_{x,y}$
states: in the $\Gamma$-A direction they are above E$_F$ and form
cylindric hole-type elements of the Fermi surface (FS), Fig. 1.\\
\begin{table}
\caption{\label{tab:table1} Diboride lattice parameters for Mg,
Zr, V, Nb, Ta by our FP LMTO self-consistent calculations and
experiment results}
\begin{ruledtabular}
\begin{tabular}{ccccccc}
Diboride&\mbox{$a$ (\AA)}&\mbox{$c$
(\AA)}&\mbox{$c/a$}&\mbox{$a^*$
(\AA)}&\mbox{$c^*$ (\AA)}&\mbox{$c/a^*$}\\\hline\\
MgB$_2$&\mbox{3.04869}&\mbox{3.46637}&\mbox{1.1370}&\mbox{3.083}&\mbox{3.521}&\mbox{1.142  [23]}\\
ZrB$_2$&\mbox{3.16932}&\mbox{3.53126}&\mbox{1.1142}&\mbox{3.170}&\mbox{3.532}&\mbox{1.114  [11]}\\
\mbox{ }&\mbox{ }&\mbox{ }&\mbox{ }&\mbox{3.165}&\mbox{3.547}&\mbox{1.120  [23]}\\
VB$_2$&\mbox{3.00678}&\mbox{3.04768}&\mbox{1.0136}&\mbox{2.997}&\mbox{3.056}&\mbox{1.0196 [15]}\\
\mbox{ }&\mbox{ }&\mbox{ }&\mbox{ }&\mbox{2.998}&\mbox{3.057}&\mbox{1.020  [23]}\\
NbB$_2$&\mbox{3.18141}&\mbox{3.35693}&\mbox{1.0550}&\mbox{3.116}&\mbox{3.264}&\mbox{1.060  [23]}\\
TaB$_2$&\mbox{3.16421}&\mbox{3.32337}&\mbox{1.0503}&\mbox{3.082}&\mbox{3.243}&\mbox{1.0522 [12]}\\
\mbox{ }&\mbox{ }&\mbox{ }&\mbox{ }&\mbox{3.098}&\mbox{3.224}&\mbox{1.0407 [14]}\\
\mbox{ }&\mbox{ }&\mbox{ }&\mbox{ }&\mbox{3.083}&\mbox{3.244}&\mbox{1.0522 [15]}\\
\mbox{ }&\mbox{ }&\mbox{ }&\mbox{ }&\mbox{3.097}&\mbox{3.225}&\mbox{1.041  [23]}\\
\end{tabular}
\end{ruledtabular}
$^*$ - [11,12,14,15,23]
\end{table}
Thus, the distinguishing characteristics of the band spectrum of
MgB$_2$, that are crucial for its superconducting properties, as
well as inter- and interlayer interaction effects (see also
[2-7]), include: (1) location of $\sigma$(p$_{x,y}$) bands
relative to E$_F$ (the presence of hole states); (2) the value of
their dispersity in the $\Gamma$-A direction
($\Delta$E$^\sigma$($\Gamma$-A), determined by the degree of
interaction between metal-boron layers); (3) the value and orbital
composition of N(E$_F$) (dominating contribution of s states of
boron atoms from graphite-like layers). Let us consider in this
context the band structure of Zr, V, Nb and Ta diborides. First
note that the most obvious consequence of the variation in the
composition of the metal sublattice in the series of diborides is
an increase in the EC with successive occupation of energy bands.
Then the Fermi level for ZrB$_2$ is located in the pseudogap
between completely occupied bonding and free antibonding states.
This determines the maximum stability of ZrB$_2$ (and also of
isoelectronic and isostructural TiB$_2$ and HfB$_2$) in the series
of AlB$_2$-like phases and their extreme thermomechanical
characteristics [23]. These inferences were confirmed by recent FP
LMTO calculations of the cohesion energy of some MB$_2$ phases (M
= 3d, 4d and 5d metals) [18, 19].\\
\begin{table}
\caption{\label{tab:table2} Dispersion of $\sigma$(P$_{x,y}$)
bands in the direction $\Gamma$-A ($\Delta$E$^\sigma$($\Gamma$-A),
eV) and orbital contributions to the density of states at the
Fermi level (states/eV, cell) in AlB$_2$ phases.}
\begin{ruledtabular}
\begin{tabular}{ccccccccc}
\mbox{ }&\mbox{$\Delta$E$^\sigma$($\Gamma$-A),}&\mbox{ }&\mbox{ }&\mbox{ }&\mbox{DOS}&\mbox{ }&\mbox{ }&\mbox{ }\\
Phase&\mbox{eV}&\mbox{Total}&\mbox{Ms}&\mbox{Mp}&\mbox{Md}&\mbox{Mf}&\mbox{Bs}&\mbox{Bp}\\\hline\\
MgB$_2$&\mbox{0.72}&\mbox{0.719}&\mbox{0.040}&\mbox{0.083}&\mbox{0.138}&\mbox{-}&\mbox{0.007}&\mbox{0.448}\\
ZrB$_2$&\mbox{1.73}&\mbox{0.163}&\mbox{0.001}&\mbox{0.002}&\mbox{0.130}&\mbox{-}&\mbox{0.000}&\mbox{0.030}\\
VB$_2$&\mbox{2.69}&\mbox{1.379}&\mbox{0.024}&\mbox{0.013}&\mbox{1.255}&\mbox{-}&\mbox{0.002}&\mbox{0.085}\\
NbB$_2$&\mbox{2.49}&\mbox{1.074}&\mbox{0.037}&\mbox{0.017}&\mbox{0.818}&\mbox{-}&\mbox{0.012}&\mbox{0.190}\\
TaB$_2$&\mbox{2.61}&\mbox{0.910}&\mbox{0.003}&\mbox{0.016}&\mbox{0.664}&\mbox{0.038}&\mbox{0.011}&\mbox{0.178}\\
\end{tabular}
\end{ruledtabular}
\end{table}
\textbf{MgB$_2$ and ZrB$_2$}. As is seen from Figs. 1, 2 and Table
2, the structure of the near-Fermi spectra edges of ZrB$_2$ and
the SC MgB$_2$ differ radically. For ZrB$_2$, (1)
$\sigma$(p$_{x,y}$) bands of boron are located in the region are
located below E$_F$ (-1.1 eV in point A BZ) and the corresponding
hole states are absent; (2) a considerable dispersion of these
bands appears in the $\Gamma$-A direction
(($\Delta$E$^\sigma$($\Gamma$-A) = 1.73 eV), $\sigma$ bands become
no longer of the 2D type as a result of formation of strong
covalent d-p bands between metal-boron layers, in which partially
occupied $\pi$(p$_z$) bands take part; (3) the value of N(E$_F$)
decreases drastically as compared with MgB$_2$ (from 0.719 to
0.163 states/eV), the maximum contribution ($\sim$80\%) to
N(E$_F$) being made by Zr4d states (whereas the contributions from
boron states are much smaller - $\sim$18\%). The change in the
type (2D $\rightarrow$ 3D) of the near-Fermi states can be easily
traced by comparing the structure of the FS of MgB$_2$ and
ZrB$_2$, Fig. 1. It is seen that the FS of ZrB$_2$ consists of
three types of figures defined by mixed Zr4d,5p-Bp states: (a) a
3D rotation figure around a straight line along the $\Gamma$-A
direction with the hole-type conductivity; (b) a 3D figure near
the centre of the M-K segment with the electronic-type
conductivity; and (c) tiny 3D-type sections with the electronic
conductivity.\\\textbf{VB$_2$, NbB$_2$ and TaB$_2$}. Energy bands,
FS and DOS of these isoelectronic and isostructural diborides are
demonstrated in Figs. 3, 4, and some electronic structure
parameters are listed in Tables 2, 3. The mentioned above
differences in ZrB$_2$ and the SC MgB$_2$ (filling of
$\sigma$(p$_{x,y}$) bands, decreased contributions of B2p states
to N(E$_F$), variation in the type (2D $\rightarrow$ 3D) of the
near-Fermi states) are typical also of VB$_2$, NbB$_2$, TaB$_2$.
Besides, they have the following features in common (as compared
with ZrB$_2$): (1) partial occupation of the antibonding d band
responsible for the metallic-type conductivity; (2) considerable
growth of N(E$_F$); and (3) increased filling of $\pi$(p$_z$)
bands. A peculiar form of the surface Fermi transformation is
observed for example for TaB$_2$ (Fig. 3): the surface Fermi
contains double ("internal" and "external") non-intersecting
electronic-type rotation spheroids around point A defined by
3D-B2p and Ta5d$_{xz,yz}$ states respectively. The bonding
$\sigma$(p$_{x,y}$) bands of boron lie in the region of -1.3, -2.5
and -2.6  eV in A-point of BZ for VB$_2$, NbB$_2$ and TaB$_2$
accordingly, below E$_F$ and have, as in the case of ZrB$_2$, a
significant energy dispersion $\Delta$E$^\sigma$($\Gamma$-A),
which has a maximum value for VB$_2$, Table 2. In the series of
isoelectronic VB$_2$ $\rightarrow$ NbB$_2$ $\rightarrow$ TaB$_2$,
N(E$_F$) decreases systematically and has a maximum value for
VB$_2$ due to the contribution from the near-Fermi quasi-plane
V3d$_{xz,yz}$ band in the direction $\Gamma$ - A. By contrast, the
contribution of B2p states (antibonding s and p bands) to N(E$_F$)
in this series changes non-monotonously and reaches its peak
(0.190) for NbB$_2$, which is much smaller than that for MgB$_2$
(0.448 states/eV). A great concentration of B2p states in the
vicinity of E$_F$ for NbB$_2$ (as compared with TaB$_2$) is also
evident from spectroscopy experiments [16]. \\
\begin{table}
\caption{\label{tab:table3} Density of states at the Fermi level
(states/eV) for Mg, Ta, V diborides according to our FP LMTO
calculations in comparison with the results [14, 15].}
\begin{ruledtabular}
\begin{tabular}{cccc}
\mbox{ }&\mbox{ }&\mbox{N$_F$}&\mbox{ }\\
Phase&\mbox{Our data}&\mbox{FP[15]}&\mbox{FPLO[14]}\\\hline\\
MgB$_2$&\mbox{0.719}&\mbox{0.691}&\mbox{0.71}\\
B2p$^*$&\mbox{0.448}&\mbox{0.494}&\mbox{-}\\\hline\\
TaB$_2$&\mbox{0.91}&\mbox{0.966}&\mbox{0.91}\\
Ta5d-&\mbox{0.664}&\mbox{0.647}&\mbox{-}\\
B2p-&\mbox{0.178}&\mbox{0.114}&\mbox{-}\\\hline\\
VB$_2$&\mbox{1.379}&\mbox{1.359}&\mbox{-}\\
V3d-&\mbox{1.255}&\mbox{1.235}&\mbox{-}\\
B2p-&\mbox{0.085}&\mbox{0.043}&\mbox{-}\\
\end{tabular}
\end{ruledtabular}
$^*$ - total DOS and partial contributions of Md and B2p states
\end{table}
Thus, the performed analysis of the band structure and the surface
Fermi of isostructural d-metal (Zr, V, Nb, Ta) diborides allows us
to formulate their fundamental differences from those of the
superconducting MgB$_2$: (1) occupation of bonding p$_{x,y}$ bands
and the absence of hole s states; (2) increased covalent
interactions between boron and metal layers (due to the
hybridization of B2p-Md states) and the loss of the energy bands
two-dimensional character; (3) changes in the value and orbital
composition of N(E$_F$), where the dominant role is played by
valence d states of metals. The latter circumstance is typical of
low-temperature superconductors, for example metal-like compounds
of these d elements with carbon, nitrogen, silicon (NbN, V$_3$Si
etc.), the T$_c$ values of which correlate with the values of
N(E$_F$) [24]. In this case the results obtained suggest that V,
Nb, Ta diborides are more likely to have low-temperature
superconductivity, and among them the maximum T$_c$ value can be
anticipated for VB$_2$. On the contrary, if we assume that the
major electronic factor for the formation of SC properties of
MB$_2$ is the near-Fermi density of B2p states (by analogy with
MgB$_2$ [2-7]), NbB$_2$ should possess the highest critical
transition temperature. It is noteworthy that according to the
coupling model proposed in [25, 26], not only boron s, but also p
bands occupation should be considered in this case. Anyhow the
superconducting transition for ZrB$_2$ is the least probable, and
the results [11] need to be revised.

\begin{figure}
\includegraphics {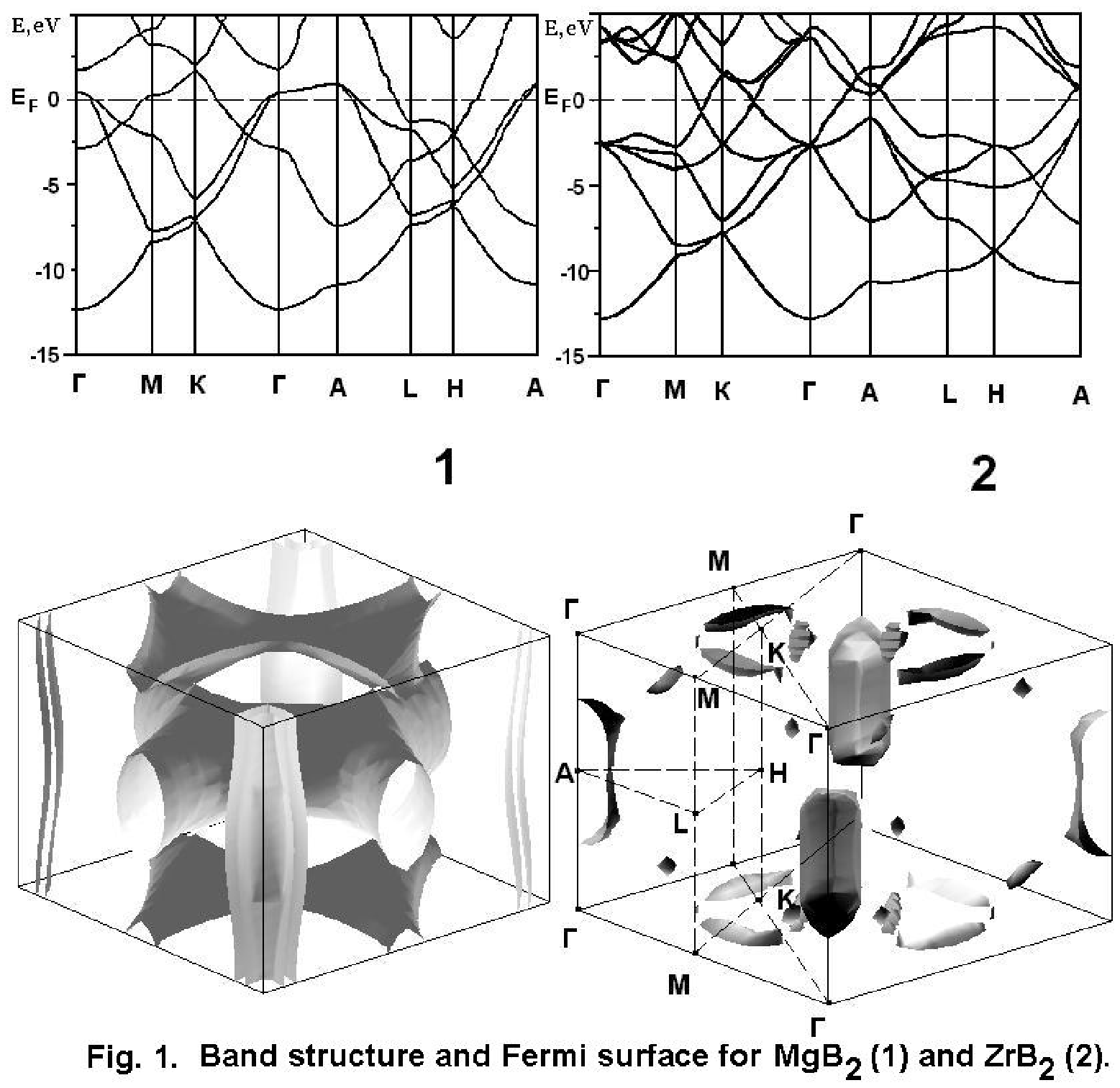}
\end{figure}

\begin{figure}
\includegraphics {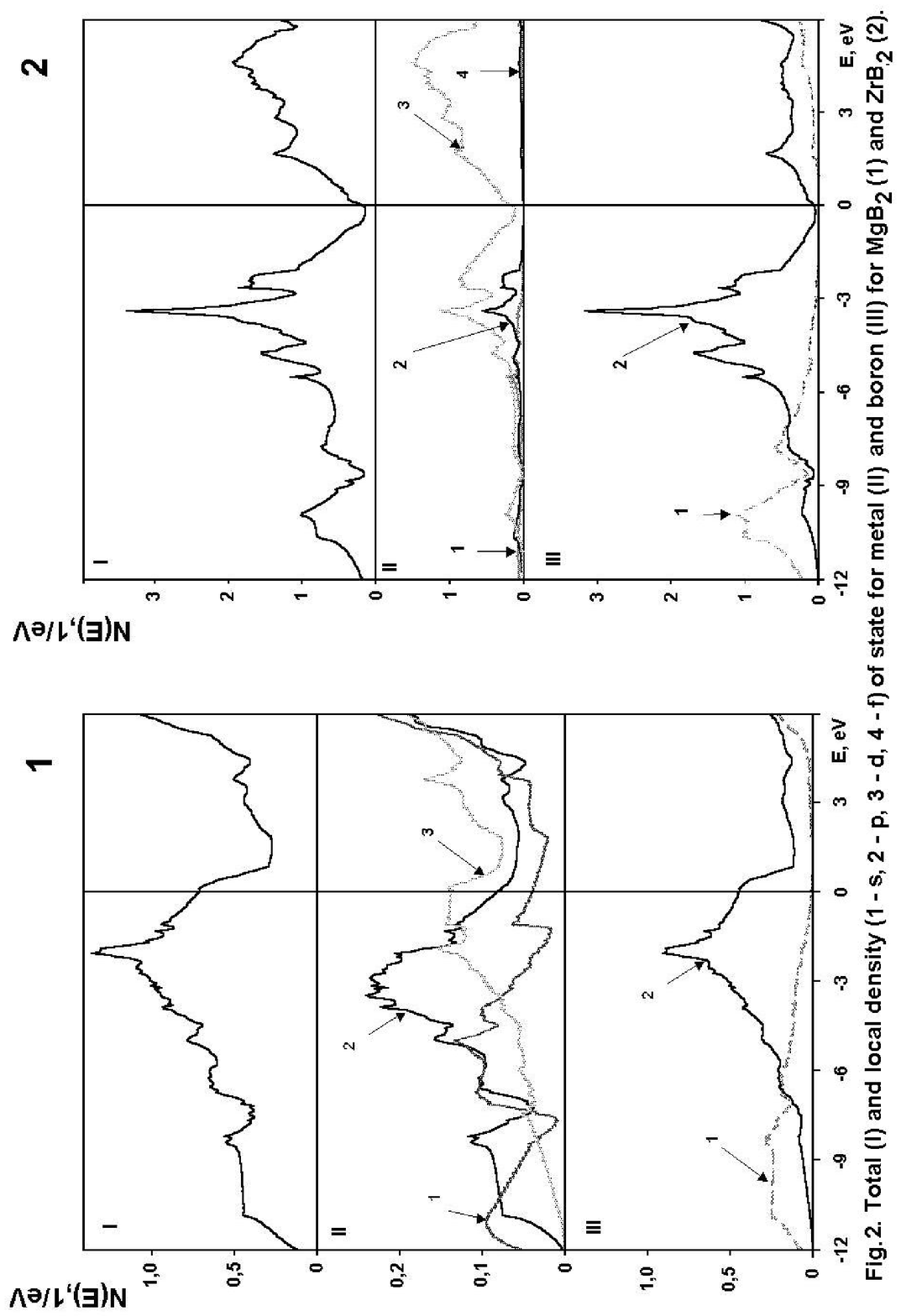}
\end{figure}

\begin{figure}
\includegraphics {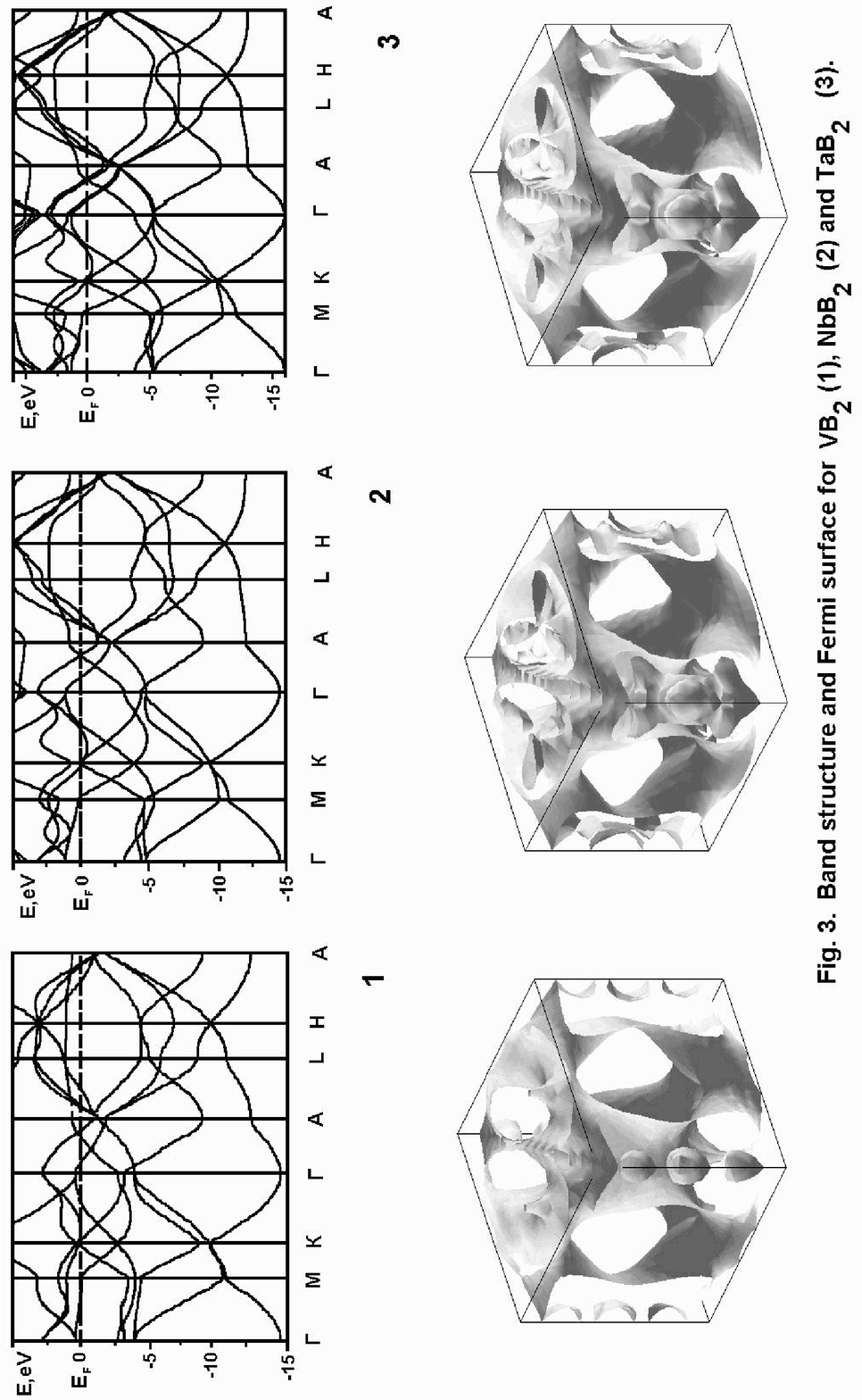}
\end{figure}

\begin{figure}
\includegraphics {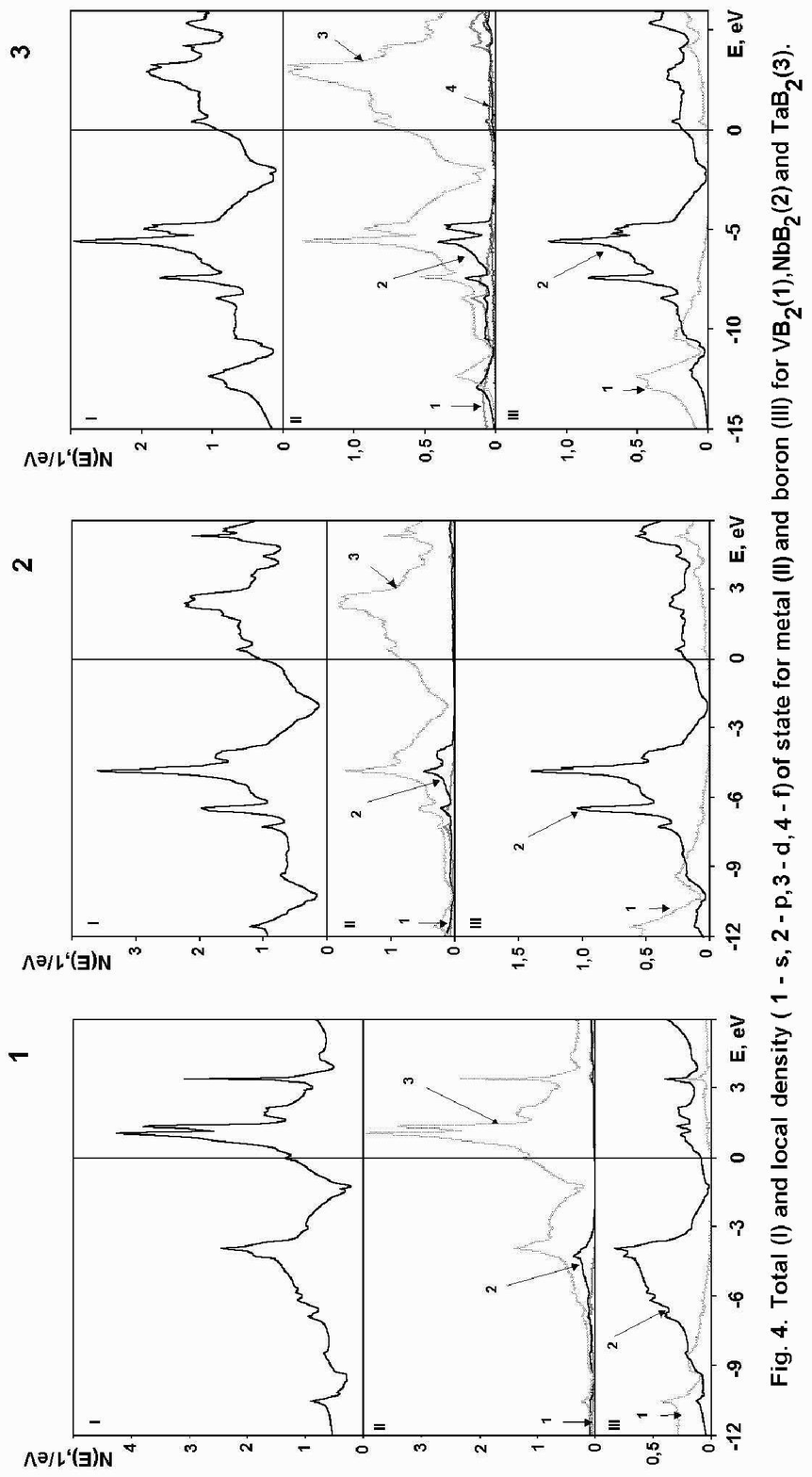}
\end{figure}

\end{document}